# Observation of Antichiral Edge States in a Circuit Lattice

Yuting Yang[1, 2, 3], Dejun Zhu[2], Zhi Hong Hang[2, 4, *] and Y. D. Chong[3, 5, †]

[1]School of Materials Science and Physics, China University of Mining and Technology, Xuzhou 221116, China

[2]School of Physical Science and Technology & Collaborative Innovation Center of Suzhou Nano Science and Technology, Soochow University, Suzhou 215006, China

[3]Division of Physics and Applied Physics, School of Physical and Mathematical Sciences, Nanyang Technological University, Singapore 637371, Singapore.

[4]Institute for Advanced Study, Soochow University, Suzhou 215006, China

[5]Centre for Disruptive Photonic Technologies, Nanyang Technological University, Singapore 637371, Singapore

*zhhang@suda.edu.cn      †yidong@ntu.edu.sg

**Abstract:**

We constructed an electrical circuit to realize a modified Haldane lattice exhibiting the phenomenon of antichiral edge states. The circuit consists of a network of inductors and capacitors with interconnections reproducing the effects of a magnetic vector potential. The next nearest neighbor hoppings are configured differently from the standard Haldane model, and as predicted by earlier theoretical studies, this gives rise to antichiral edge states that propagate in the same direction on opposite edges and co-exist with bulk states at the same frequency. Using pickup coils to measure voltage distributions in the circuit, we experimentally verify the key features of the antichiral edge states, including their group velocities and ability to propagate consistently in a Möbius strip configuration.

The Haldane model[1] is a simple but rich theoretical model that exemplifies the physics of two-dimensional topological insulators[2-5]. When magnetic fluxes are appropriately threaded through a honeycomb lattice, with zero net flux in each unit cell, the band structure hosts a band gap spanned by chiral edge states; on a rectangular strip, the edge states localized on opposite edges will propagate in opposite directions. The edge states are protected by topological band invariants (Chern numbers) of the bulk bands. The Haldane model has been realized in condensed matter systems[6], and very similar models have been used to create classical wave analogues of topological insulators based on electric circuits[7-11], photonics[12-14] and acoustics[15-17]. The Haldane model has also inspired the development of more complex topological insulators, such as the time-reversal (T) invariant Kane-Mele model[18,19], which consists of spin-up and spin-down sectors that can be regarded as two copies of the Haldane model.

Recently, Colomés and Franz discovered that a subtle modification to the Haldane model leads to a strikingly different behavior[20]. With a different configuration of magnetic fluxes, equivalent to reversing the next-nearest-neighbor (NNN) hoppings in one sublattice, the lattice exhibits "antichiral edge states" that propagate in the same direction on opposite sides of a rectangular strip. Moreover, the bulk spectrum is ungapped, so on a finite rectangular strip the transmission in one direction is edge-dominated whereas transmission in the opposite direction must occur via the bulk[20]. To our knowledge, there is thus far no experimental demonstration of this effect, despite proposals to realize it using strained materials[21], ferromagnetic materials with Dzyaloshinskii-Moriya interactions[22], exciton polaritons[23], gyromagnetic photonic crystals[24], and graphene[25,26].

Here, we use electric circuits to experimentally realize antichiral edge states and study their properties. Circuit metamaterials have been the subject of recent theoretical and experimental interest[27-39] due to the ease with which they can be designed and fabricated to realize different topological phases, as well as unusual lattice configurations that are hard to achieve on other platforms. Circuits have been used to demonstrate nonlinear topological boundary states[32,33], topological corner modes[34-37] and four-dimensional topological insulators[38,39]. Most notably, Jia *et al.* have shown how a Haldane-type Chern insulator phase can be accessed using a lattice of capacitors (C) and inductors (L) with braided interconnections[7]. Although LC circuits are time-reversal symmetric, the braiding decomposes the spectrum into two degenerate decoupled sectors that are individually T-broken[27-42], with the physical T symmetry mapping each sector to the other. Utilizing this idea, we design and fabricate a braided LC circuit lattice that realizes the modified Haldane model. Using different samples with electrical connections simulating periodic or closed boundaries, we probe the bulk and edge excitation spectra as well as the antichiral propagation characteristics of the edge states, which agree well with theoretical predictions. Moreover, we show experimentally that the antichiral states are able to propagate consistently (i.e., without switching sectors) along the edge of a Möbius strip edge – a property that the chiral edge states of a Chern insulator does not have[7]. This work points the way toward using circuit metamaterials for future studies of more complicated T-broken materials, including higher dimensional lattices and unusual sample geometries.

The schematic of the modified Haldane model[20] is shown in Fig. 1**a**. Each unit cell contains two sites, A and B. The NNN hoppings between A sites and between B sites have π/2 phase shifts in the directions indicated by the arrows. The nearest-neighbor (NN) hoppings have zero phase. Figure 1**b** shows schematically how such hoppings can be realized using interconnected capacitor and inductor elements. On each lattice site there are two inductors $X$ and $Y$, whose ends are labeled as $X_\pm$ and $Y_\pm$; the voltages across the inductors are $U_X = V_{X_+} - V_{X_-}$ and $U_Y = V_{Y_+} - V_{Y_-}$ respectively. All the inductors have the same inductance $L$, and inductors at different sites are connected via capacitors. For NN (zero phase) hoppings, we connect each end $X_\pm$ to $Y_\pm$ with capacitors of capacitance $C_1$. For NNN (π/2 phase) hoppings, we use capacitors of capacitance $C_2$, and the connections are braided so that $U_X \to -U_Y$ and $U_Y \to U_X$. Defining $U_{\uparrow,\downarrow} = U_X \pm iU_Y$, we find that the NNN hoppings correspond to $U_\uparrow \to iU_\uparrow$ and $U_\downarrow \to -iU_\downarrow$, as desired[7,42]. Henceforth, we will focus on one of the two "spin" sectors, specifically the up spin.

For steady-state solutions of angular frequency ω, we can show using Kirchhoff's laws (see Supplementary Information) that

$$E \begin{pmatrix} U^A_{k,\uparrow} \\ U^B_{k,\uparrow} \end{pmatrix} = \begin{pmatrix} P_k(\phi) & T_k \\ T_k^* & P_k(-\phi) \end{pmatrix} \begin{pmatrix} U^A_{k,\uparrow} \\ U^B_{k,\uparrow} \end{pmatrix}, \quad (1)$$

where $E = 3t_1 + 6t_2 - 2\omega_0^2/\omega^2$, $\omega_0 = 1/\sqrt{LC}$, and $C$ is a reference capacitance such that $C_1 = t_1 C$ and $C_2 = t_2 C$. This has the form of the modified Haldane model, with the caveat that the eigenvalue $E$ is not equal to the eigenfrequency. The Hamiltonian matrix elements are defined by $T_k = t_1\left(e^{i k \vec{e}_1} + e^{i k \vec{e}_2} + e^{i k \vec{e}_3}\right)$ and $P_k = 2t_2\left[\cos(k\vec{v}_1 + \phi) + \cos(k\vec{v}_2 + \phi) + \cos(k\vec{v}_3 + \phi)\right]$, where $\phi = \pi/2$ is the NNN hopping phase. As indicated in the right panel of Fig. 1a, $\vec{e}_i$ ($i$=1, 2, 3) are the NN bond vectors, and $\vec{v}_j$ ($j$=1, 2, 3) are the NNN bond vectors.

We choose the circuit parameters to be $L$=3.3 mH, $C_1$ =330 pF and $C_2$ =33 pF, so that the eigenfrequency is related to $E$ by $f = \dfrac{1}{2\pi}\dfrac{1}{\sqrt{LC}}\sqrt{\dfrac{2}{3.6-E}}$ where $C = 330$ pF (i.e., $t_1 = 1$ and $t_2 = 0.1$). Figure 1c shows the resulting band diagram, with the physical eigenfrequency $f$ as the vertical axis, for a strip that is infinite along $x$ and 20 unit cells wide along $y$, with zigzag boundaries. In agreement with the prior findings of Colomés and Franz[20], the Dirac points are shifted in opposite directions, and joined by a two-fold degenerate arc.

Figure 1d shows the intensity distributions for four of the eigenstates at wavenumber $k = \pi/a$. The middle two panels, labelled $u_2$ and $u_3$, correspond to the two degenerate eigenmodes at frequency 113.63 kHz (red dashes in Fig. 1c); they are localized to opposite edges of the strip, despite having the same group velocity as shown in Fig. 1c. The other eigenstates are bulk states, as exemplified by the eigenstates labelled $u_1$ and $u_4$, which occur at frequencies 105.08 and 124.28 kHz (blue dashes in Fig. 1c). The results shown here are for the spin-up states. For the spin-down states, the antichiral edge states have the opposite group velocity (Supplementary Fig. S1).

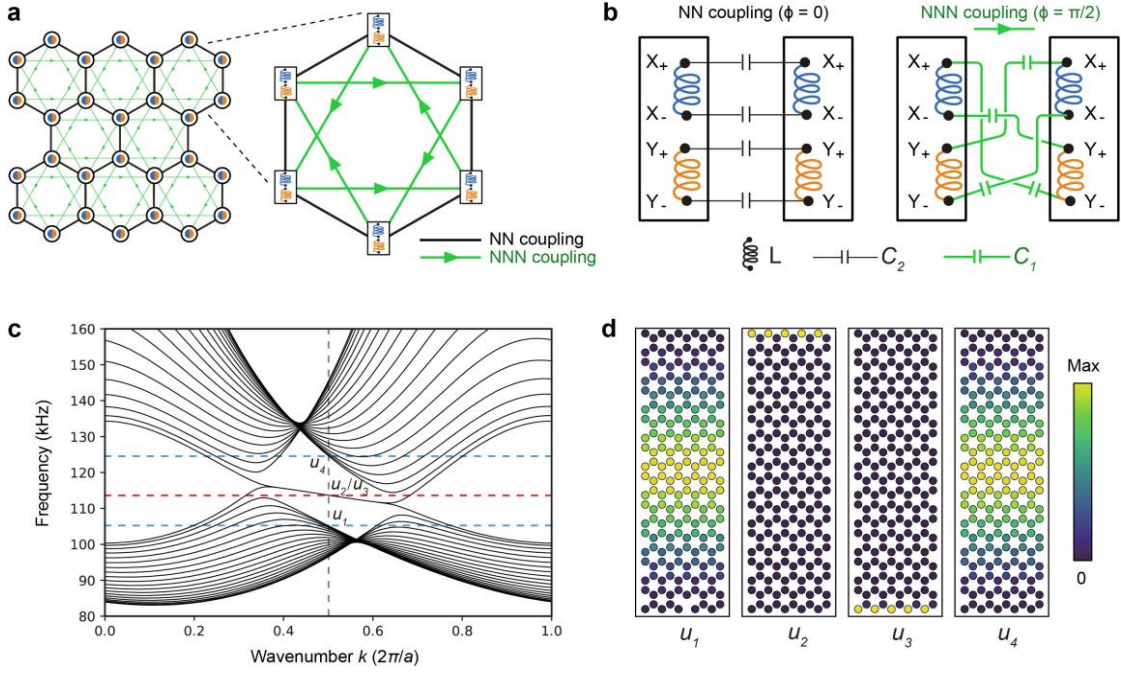

**Fig. 1. Setup of the circuit. a** Schematic of a modified Haldane mode. Each unit cell consists of sites A and B, with each site containing a pair of inductors. NN and NNN hoppings are indicated by black lines and green arrows. **b** Schematic of the braided capacitive couplings between inductors on different sites for NN hoppings (with zero phase) and NNN hoppings (with phase $\pi/2$) respectively. **c** Band diagram of a semi-infinite circuit lattice consisting of a strip 20 unit cells wide and infinitely long, with zigzag boundaries and circuit parameters $L$=3.3 mH, $C_1$ =330 pF, and $C_2$ =33 pF. **d** Intensity distributions at wavenumber $\pi/a$ for the two degenerate states at frequency 113.63 kHz ($u_2$ and $u_3$) and two representative bulk states at 105.08 and 124.28 kHz ($u_1$ and $u_4$). The antichiral edge states are localized to different edges but have the same group velocity.

## Results

We implemented the lattice using a FR4 printed circuit board (PCB), as shown in Fig. 2**a**. The black cylinders in the photograph are wire-wound inductors, and the yellow components are coupling capacitors. The circuit parameters are as stated in the previous paragraph. The PCB contains additional traces that can be used to connect additional inductors. The PCB is two sites wide in the *x* direction and 10 sites (5 unit cells) wide in the *y* direction; the site numbers are explicitly labeled (1 to 20) in the photograph.

We first connect the left and right boundaries using capacitive connections in order to realize periodic boundary conditions along *x*. Since the circuit is two unit cells wide along *x*, this is equivalent to probing $k = 0$ and $k = \pi/a$ in the band diagram of Fig. 1**a**, with the latter allowing the antichiral edge states to be accessed. Additional shorted-out capacitors are added to the top and bottom *y* boundaries to achieve clean zigzag boundaries, compensating for the change in on-site potential caused by missing couplings at the lattice terminations. We place driving coils on the X inductors at sites 2 and 12, and use a pickup coil to measure the voltage amplitude on the Y inductor at site 1 (see Methods). The results are shown in the red curve in Fig. 2**b**. The response is peaked at 111.1 kHz, close to the predicted frequency of the antichiral edge states shown in Fig. 1c. Although the antichiral edge states coexist with bulk states, they are preferentially excited due to the strong spatial overlap with the driving source. Note that this driving scheme excites both spin-up and spin-down states, but the antichiral edge states in both spin sectors are degenerate at $k = \pi/a$. These experimental findings agree well with the results of circuit simulations (see Methods), shown by the red curve in Fig. 2c. In the simulations, the response peaks at 113.64 kHz. This small frequency shift can be attributed to fabrication errors, such as the approximately 5% tolerance in the capacitances and inductances of the various circuit components.

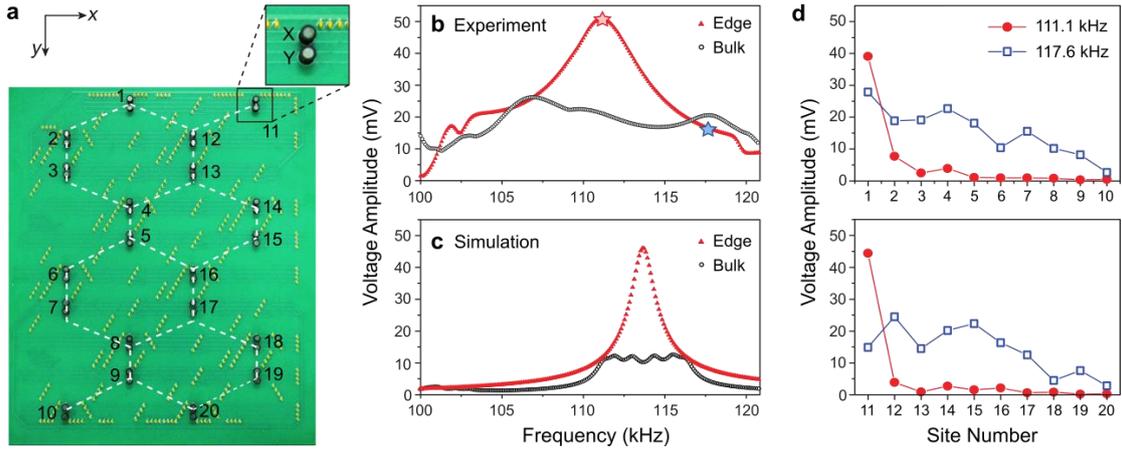

**Fig. 2. Experimental characterization of bulk and edge states. a** Photograph of a circuit corresponding to a lattice with periodic boundaries in $x$ and zigzag boundaries in $y$. The 20 sites in the sample are explicitly numbered. Each site has an X and Y inductor (black cylinders), with NN and NNN hoppings implemented via capacitors (yellow components). **b** Experimentally measured voltage amplitudes. For the red curve, driving coils are placed on the X inductors at sites 2 and 12, and the pickup coil is placed on the Y inductor at site 1; the excitation and measurement thus occur along one edge of the effectively semi-infinite strip, and the peak at 111.1 kHz (red star) is close to the predicted eigenfrequency of the antichiral edge states. For the black curve, driving coils are placed on the X inductors at sites 1 and 4/6/8 (results are averaged over the three driving configurations), and the pickup coil is placed on the Y inductor at site 17; this serves as a probe of the spatially-averaged density of states, and the lack of a dip in the response shows the lack of a bulk band gap. **c** Circuit simulation results corresponding to Fig. 2b. For the strip geometry, the response peaks at frequency 113.64 kHz, close to the experimental peak. **d** Voltage amplitudes measured at different sites, showing strong edge localization at 111.1 kHz (red), and no edge localization at frequency 117.6 kHz (blue). The driving coils are placed on the X inductors at sites 2 and 12, and the pickup coils are placed on the Y inductors at different sites.

Next, we studied the bulk lattice by connecting the top and bottom inductors of the strip and removing the additional shorted-out capacitors, which is equivalent to applying periodic boundary conditions to opposite edges of the strip. To probe the spatially averaged density of states, we excite the lattice using one driving source at site 1 and another at sites 4, 6, or 8, on the X inductors, with the pickup coil located at the X inductor on site 17, and take the averages of the three data sets. The results, plotted as the black curve in Fig. 2**b**, show no significant dip in the frequency range of interest. This agrees with the theoretical expectation that this bulk bandstructure, unlike that of the standard Haldane model, lacks a band gap. The circuit simulation results, shown as the black curve in Fig. 2**c**, exhibit qualitatively similar behavior.

The localized nature of the edge states can be observed by taking voltage amplitude measurements at different sites. Figure 2**d** shows the experimentally measured voltage amplitudes at the Y inductors on different sites, for the previously-discussed strip geometry (i.e., open boundary conditions along the edges of the strip, with driving coils on the X inductors at sites 2 and 12, corresponding to the red curve in Fig. 2**b**). Here, the red curves show the response at the peak frequency of 111.1 kHz, which is strongly localized the top edge (the edge states on the bottom edge are not excited since the sources are located on the top edge). For comparison, the blue curves show the response at 117.6 kHz, away from the eigenfrequency of the edge states, for which the response is not localized to the edge.

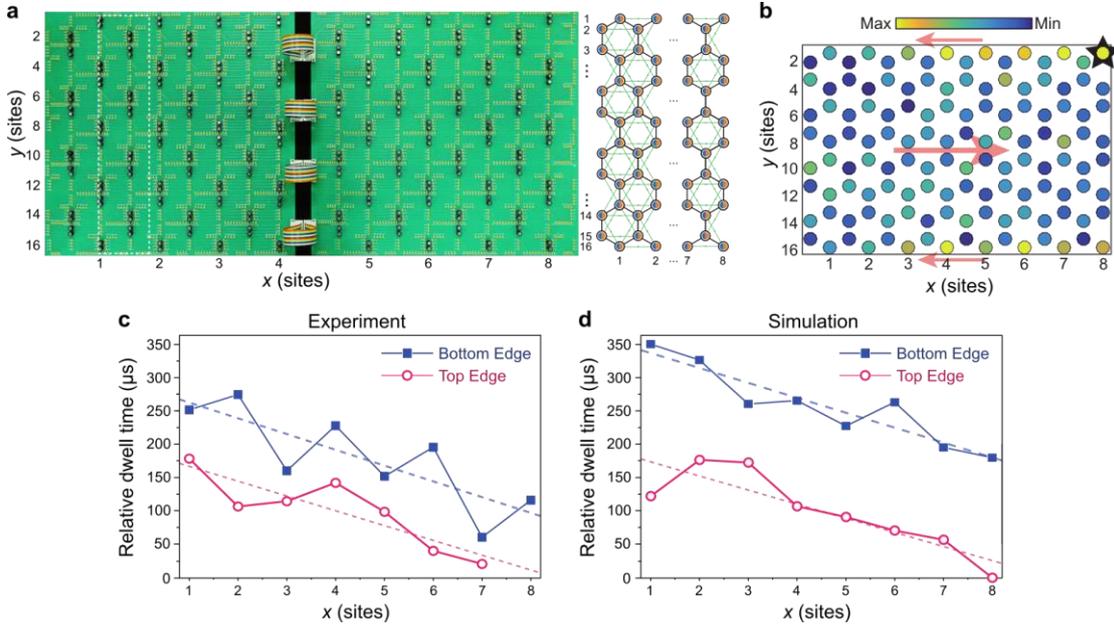

**Fig. 3. Propagation of antichiral edge states in a finite lattice. a** Photograph of the circuit boards implementing a 128-site lattice with open boundary conditions (left panel), and the schematic of the lattice (right panel). **b** Experimental results showing the voltage amplitude distribution at 112.7 kHz, produced by two driven coils with 90° phase difference (so as to excite spin-up states) placed on inductors X and Y at a corner site (marked by a black star). Red arrows indicate the expected propagation directions of the edge and bulk states. **c** Experimentally measured relative dwell times along the top edge (pink) and bottom edge (blue). The slopes of the linear least-squares fits correspond to group velocities of -0.045 sites/μs (top edge) and -0.042 sites/μs (bottom edge). **d** Relative dwell times obtained from corresponding circuit simulations. The slopes of the linear least-squares fits correspond to group velocities of -0.047 sites/μs (top edge) and -0.045 sites/μs (bottom edge).

To further characterize the antichiral edge states, we prepared a circuit corresponding to a finite lattice of 128 sites in a rectangular geometry (see Fig. 3**a**). Owing to fabrication limitations, the sample consists of two PCBs connected by cable assemblies. We apply two driving coils to the X and Y inductors at a corner site (marked by a black star in Fig. 3**b**), with a 90° relative phase shift in order to selectively excite spin-up states. In this configuration, the driving coils should excite antichiral edge states that propagate leftward along the upper edge. Figure 3**b** shows the experimentally obtained spin-up voltage amplitude distribution at 112.7 kHz (the frequency matching the antichiral edge states at $k = \pi/a$, as seen in Fig 2**b**). A strong voltage response is observed at both sample edges, a result that agrees well with the steady-state voltage distributions obtained in circuit simulations (Supplementary Fig. S3**a**). These results are consistent with the interpretation that antichiral edge states are initially excited on the top edge, undergo reflection at the left boundary into the bulk states, and reflect off the right boundary into left-moving antichiral edge states on both edges[20]. This behavior is further confirmed by time-domain circuit simulations (Supplementary Fig. S4).

We then determine the group velocities of the edge states by measuring the dwell time $d\varphi/d\omega$, where $\varphi$ is the phase of the complex spin-up voltage measured by the pickup coils at each site, and $\omega$ is the angular frequency. The rate of change of dwell time with distance along the edge is the group velocity[43]. Figure 3**c** shows the experimental results for the dwell times on the top and bottom edges; each data point is estimated from spin-up voltage measurements at 5 equally-spaced frequencies between 112.5 kHz and 112.9 kHz (Supplementary Fig. S3**b**). From a linear least-squares fit of these results, we estimate group velocities of -0.045 sites/μs (top edge) and -0.042 sites/μs (bottom edge). The group velocities on both edges are negative, consistent with theoretical predictions. The corresponding circuit simulations (see Fig. 3**d**) predict group velocities of -0.047 sites/μs (top edge) and -0.045 sites/μs (bottom edge). From the band diagram of a strip of the same width and infinite length (similar to Fig. 1**c** but with reduced width), the group velocities of the antichiral edge states is estimated to be -0.059 sites/μs. These experimental results unambiguously verify the antichiral nature of the edge states.

One of the most interesting features of electrical circuits is the ability to set up lattice geometries that are difficult or impossible to realize on other platforms. Jia *et al.*, for instance, showed that a rectangular sample can be converted into a Möbius strip by placing appropriate electrical connections between sites on two opposite boundaries[7]. However, an unpaired Chern insulator cannot be placed in a Möbius strip geometry – an edge state, upon crossing one boundary, passes through to the opposite edge of the strip moving in the same direction, which is inconsistent with chiral propagation. In an actual circuit, the edge states switch to the opposite spin upon crossing the boundary[7]; in other words, the Möbius strip geometry necessarily couples the two spin sectors. The modified Haldane lattice, however, *can* be self-consistently implemented on a Möbius strip since its edge states are antichiral. To test this notion, we implemented a circuit with a Möbius strip configuration via twisted connections between sites on opposite boundaries, as shown in Fig. 4**a**. As shown in Figs. 4**b** and **c**, the antichiral edge states are able to traverse the entire Möbius strip edge, traveling in the same direction along the top and bottom edges. These results are consistent with circuit simulations (Supplementary Fig. S5).

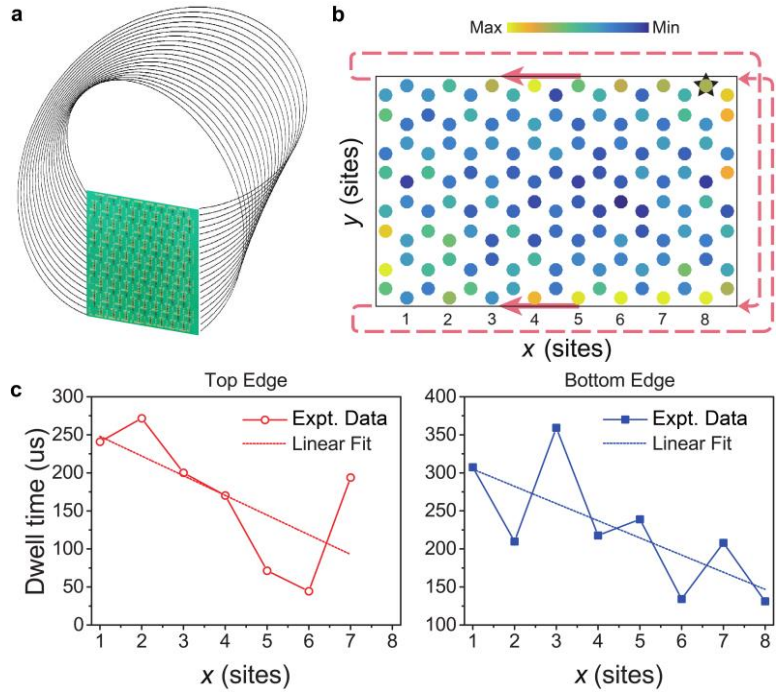

**Fig. 4. Propagation of antichiral edge states in a Möbius strip. a** Schematic of the Möbius strip circuit. Twisted electrical connections are applied to the left and right boundaries of the physical sample, so that the upper edge of the strip continues to the lower edge and vice versa. **b** Experimental results showing the voltage amplitude distribution at 116.4 kHz with spin up excitation. Red arrows indicate the propagation directions of the edge and bulk states. **c** Experimentally measured relative dwell times along the top edge (red) and bottom edge (blue). The slopes of the linear least-squares fits correspond to group velocities of -0.039 sites/μs (top edge) and -0.044 sites/μs (bottom edge).

**Discussion**

We have experimentally verified the key features of the modified Haldane lattice proposed by Colomés and Franz[20], including the existence of antichiral edge states that have the same group velocity on opposite edges, the lack of a bulk gap, the transfer of energy between edge and bulk states during successive reflections within a finite sample, and the ability to exist self-consistently on a Möbius strip. These results demonstrate the flexibility of electrical circuits, as an experimental platform, for realizing topological phases and other related lattice phenomena[7,27]. In particular, we have used the "braiding" trick, originally used by Jia *et al.* to realize a Chern insulator[7], to implement a new type of effective vector potentials (complex inter-site hoppings). In the future, this approach might be used to implement circuit lattices with even more complicated vector potentials. Bulk bandstructure features, such as the Berry curvature, can also be probed using previously-established tomographic methods[40,44]. It would be also interesting to further investigate the use of circuit lattices to implement nontrivial vector potentials in other non-traditional geometries such as Klein bottles.

## Methods

**Experimental setup:** All circuits were implemented on FR4 printed circuit boards. Components consist of unshielded wire-wound inductors with 13 Ω series resistance, and ceramic capacitors. The input signal is produced by a function generator (Tektronix AFG3022C) connected to 9-turn, 8-mm-diameter air-core driving coils. The output signal is obtained with a pickup coil of the same dimensions, connected to a lock-in amplifier (Zurich Instrument MFLI).

**Circuit simulations:** Circuit simulations were performed using the LTspice circuit simulator. Inductors are given series resistance 13 Ω, consistent with the inductors on the PCB. AC (steady state) analysis is used for the simulations shown in the main text, with a 1V sine wave as the source. The voltage at each inductor node is probed, and the amplitude and phase are used to derive the complex signal.

## Data availability

Raw experimental and simulated data for all figures can be found at https://doi.org/10.21979/N9/CLFQXH. The other data that support the findings of this study in the paper are available from the corresponding author upon reasonable request.

## References

1. Haldane, F. D. M. Model for a quantum Hall effect without Landau levels: Condensed-matter realization of the "parity anomaly". Phys. Rev. Lett. **61**, 2015 (1988).
2. Hasan, M. Z. & Kane, C. L. Colloquium: Topological insulators. Rev. Mod. Phys. **82**, 3045 (2010).
3. Qi, X.-L. & Zhang, S.-C. Topological insulators and superconductors. Rev. Mod. Phys. **83**, 1057–1110 (2011).
4. Lu, L., Joannopoulos, J. D. & Soljačić, M. Topological photonics. Nat. Photonics **8**, 821-829 (2014).
5. Khanikaev, A. B. & Shvets, G. Two-dimensional topological photonics. Nat. Photonics **11**, 763-773 (2017).
6. Jotzu, G. et al. Experimental realization of the topological Haldane model with ultracold fermions. Nature **515**, 237-240 (2014).
7. Jia, N., Owens, C., Sommer, A., Schuster, D., & Simon, J. Time-and site-resolved dynamics in a topological circuit. Phys. Rev. X **5**, 021031 (2015).
8. Albert, V. V., Glazman, L. I., & Jiang, L. Topological properties of linear circuit lattices. Phys. Rev. Lett. **114**, 173902 (2015).
9. Hofmann, T., Helbig, T., Lee, C. H., Greiter, M., & Thomale, R. Chiral voltage propagation and calibration in a topolectrical Chern circuit. Phys. Rev. Lett. **122**, 247702 (2019).
10. Zhang, Z.-Q., Wu, B.-L., Song, J. & Jiang, H. Topological Anderson insulator in electric circuits. Phys. Rev. B **100**, 184202 (2019).
11. Ezawa, M. Electric circuits for non-Hermitian Chern insulators. Phys. Rev. B **100**, 081401(R) (2019).
12. Raghu, S. & Haldane, F. D. M. Analogs of quantum-Hall-effect edge states in

photonic crystals. Phys. Rev. A **78**, 033834 (2008).

13. Wang, Z., Chong, Y. D., Joannopoulos, J. D. & Soljačić, M. Observation of unidirectional backscattering-immune topological electromagnetic states. Nature **461**, 772-775 (2009).
14. Poo, Y., Wu, R. -X., Lin, Z., Yang, Y. & Chan, C. T. Experimental realization of self-guiding unidirectional electromagnetic edge states. Phys. Rev. Lett. **106**, 093903 (2011).
15. Fleury, R., Sounas, D. L., Sieck, C. F., Haberman, M. R. & Alù, A. Sound isolation and giant linear nonreciprocity in a compact acoustic circulator. Science **343**, 516–519 (2014).
16. Yang, Z. et al. Topological acoustics. Phys. Rev. Lett. **114**, 114301 (2015).
17. Xiao, M., Chen, W.-J., He, W.-Y. & Chan, C. T. Synthetic gauge flux and Weyl points in acoustic systems. Nat. Phys. **11**, 920-924 (2015).
18. Kane, C. L. & Mele, E. J. Quantum spin Hall effect in graphene. Phys. Rev. Lett. **95**, 226801 (2005).
19. Bernevig, B. A., Hughes, T. L. & Zhang, S. C. Quantum spin Hall effect and topological phase transition in HgTe quantum wells. Science **314**, 1757 (2006).
20. Colomés, E. & Franz, M. Antichiral edge states in a modified Haldane nanoribbon. Phys. Rev. Lett. **120**, 086603, (2018).
21. Mannaï, M., & Haddad, S. Strain tuned topology in the Haldane and the modified Haldane models. J. Phys. Condens. Matter **32**, 225501 (2020).
22. Bhowmick, D., & Sengupta, P. Antichiral edge states in Heisenberg ferromagnet on a honeycomb lattice. Phys. Rev. B, **101**, 195133 (2020).
23. Mandal, S., Ge, R., & Liew, T. C. H. Antichiral edge states in an exciton polariton strip. Phys. Rev. B **99**, 115423 (2019).
24. Chen, J., Liang, W. & Li, Z.-Y. Antichiral one-way edge states in a gyromagnetic photonic crystal. Phys. Rev. B **101**, 214102 (2020).
25. Wang, C., Zhang, L., Zhang, P., Song, J., & Li, Y.-X. Influence of antichiral edge states on Andreev reflection in graphene-superconductor junction. Phys. Rev. B **101**, 045407 (2020).
26. Denner, M. M., Lado, J. L., & Zilberberg, O. Anti-chiral states in twisted graphene multilayers. arXiv preprint arXiv:2006.13903 (2020).
27. Lee, C. H. et al. Topolectrical circuits. Comm. Phys. 1, **39** (2018).
28. Goren, T., Plekhanov, K., Appas, F. & Le Hur, K. Topological Zak phase in strongly coupled LC circuits. Phys. Rev. B **97**, 041106 (R) (2018).
29. Luo, K., Feng, J., Zhao, Y. X., & Yu R. Nodal manifolds bounded by exceptional points on non-Hermitian honeycomb lattices and electrical-circuit realizations. arXiv preprint arXiv:1810. 09231 (2018).
30. Luo, K., Yu, R., & Weng, H. Topological nodal states in circuit lattice. Research **2018**, 6793752 (2018).
31. Helbig, T. et al. Band structure engineering and reconstruction in electric circuit networks. Phys. Rev. B **99**, 161114(R) (2019).
32. Hadad, Y., Soric, J. C., Khanikaev, A. B., & Alù, A. Self-induced topological protection in nonlinear circuit arrays. Nat. Electron. **1**, 178-182 (2018).

33. Wang, Y., Lang, L. J., Lee, C. H., Zhang, B. & Chong, Y. D. Topologically enhanced harmonic generation in a nonlinear transmission line metamaterial. Nat. Commun. **10**, 1102 (2019).
34. Ezawa, M. Higher-order topological electric circuits and topological corner resonance on the breathing kagome and pyrochlore lattices. Phys. Rev. B, **98**, 201402 (2018).
35. Imhof, S. et al. Topolectrical-circuit realization of topological corner modes. Nat. Phys. **14**, 925-929 (2018).
36. Serra-Garcia, M., Süsstrunk, R., & Huber, S. D. Observation of quadrupole transitions and edge mode topology in an LC circuit network. Phys. Rev. B **99** 020304(R) (2019).
37. Yang, H., Li, Z.-X., Liu Y., Cao Y., & Yan, P. Observation of symmetry-protected zero modes in topolectrical circuits. Phys. Rev. Research **2**, 022028 (R) (2020).
38. Yu, R., Zhao, Y. X., & Schnyder, A. P. 4D spinless topological insulator in a periodic electric circuit. Natl. Sci. Rev. nwaa065 (2020).
39. Wang, Y., Price, H. M., Zhang, B., & Chong, Y. D. Circuit implementation of a four-dimensional topological insulator. Nat. Comm. **11**, 2356 (2020).
40. Lu, Y. et al. Probing the Berry curvature and Fermi arcs of a Weyl circuit. Phys. Rev. B **99**, 020302 (2019).
41. Zhu, W., Hou, S., Long, Y., Chen, H., & Ren, J. Simulating quantum spin Hall effect in the topological Lieb lattice of a linear circuit network. Phys. Rev. B **97**, 075310 (2018).
42. Zhu, W., Long, Y., Chen, H., & Ren, J. Quantum valley Hall effects and spin-valley locking in topological Kane-Mele circuit networks. Phys. Rev. B **99**, 115410 (2019).
43. Cheng, X. et al. Robust reconfigurable electromagnetic pathways within a photonic topological insulator. Nat. Mater. **15**, 542–548 (2016).
44. Owens, C. et al. Quarter-flux Hofstadter lattice in a qubit-compatible microwave cavity array. Phys. Rev. A **97**, 013818 (2018).

**Acknowledgements**
We thank You Wang, Qiang Wang and Udvas Chattopadhyay at Nanyang Technological University for helpful discussions. This work is supported by National Natural Science Foundation of China (No. 11874274), Natural Science Foundation of Jiangsu Province (No. BK20170058 and No. BK20200630), and a Project Funded by the Priority Academic Program Development of Jiangsu Higher Education Institutions (PAPD). Y. D. C. was supported by the Singapore MOE Academic Research Fund Tier 3 Grant MOE2016-T3-1-006.

**Author contributions**
Y. Y. performed the theoretical calculations and simulations, designed the printed circuit boards. D. Z. performed the experiments and Y. Y. analyzed the experimental data. Y. Y., Z. H. H. and Y. D. C. contributed to the data analysis and wrote the paper. Z. H. H. and Y. D. C. supervised the project. All authors contributed substantially to the work.

**Competing interests**
The authors declare no competing interests.

# Supplementary Information for

# Observation of Antichiral Edge States in a Circuit Lattice

Yuting Yang, Dejun Zhu, Zhi Hong Hang, and Y. D. Chong

## Circuit Equations

The circuit consists of inductors laid out in a honeycomb lattice, and A and B sites each containing a pair of inductors. The sites are labeled by indices ($m$, $n$), as shown in Fig. S1**a**. The ends of the inductors are connected to those of the inductors on other sites via capacitors, as shown in Fig. 1**b** of the main text. By using Kirchhoff's laws, the circuit equation of one node $X^+$ of inductor X at site ($m$, $n$, A) can be written as

$$\frac{V^A_{m,n,X^+} - V^A_{m,n,X^-}}{2i\omega L} + i\omega t_1 C(V^A_{m,n,X^+} - V^B_{m,n,X^+}) + i\omega t_1 C(V^A_{m,n,X^+} - V^B_{m,n-1,X^+}) + i\omega t_1 C(V^A_{m,n,X^+} - V^B_{m+1,n-1,X^+})$$
$$+ i\omega t_2 C[(V^A_{m,n,X^+} - V^A_{m,n+1,Y^-}) + (V^A_{m,n,X^+} + V^A_{m,n-1,Y^-}) + (V^A_{m,n,X^+} + V^A_{m+1,n,Y^-})$$
$$+ (V^A_{m,n,X^+} - V^A_{m-1,n,Y^-}) + (V^A_{m,n,X^+} + V^A_{m-1,n+1,Y^-}) + (V^A_{m,n,X^+} - V^A_{m+1,n-1,Y^-})] = 0$$

. (1)

And for the other node $X_-$,

$$\frac{V^A_{m,n,X^-} - V^A_{m,n,X^+}}{2i\omega L} + i\omega t_1 C(V^A_{m,n,X^-} - V^B_{m,n,X^-}) + i\omega t_1 C(V^A_{m,n,X^-} - V^B_{m,n-1,X^-}) + i\omega t_1 C(V^A_{m,n,X^-} - V^B_{m+1,n-1,X^-})$$
$$+ i\omega t_2 C[(V^A_{m,n,X^-} - V^A_{m,n+1,Y^+}) + (V^A_{m,n,X^-} + V^A_{m,n-1,Y^+}) + (V^A_{m,n,X^-} + V^A_{m+1,n,Y^+})$$
$$+ (V^A_{m,n,X^-} - V^A_{m-1,n,Y^+}) + (V^A_{m,n,X^-} + V^A_{m-1,n+1,Y^+}) + (V^A_{m,n,X^-} - V^A_{m+1,n-1,Y^+})] = 0$$

. (2)

The difference between Eq. (1) and Eq. (2) yields

$$U^A_{m,n,X} = -\frac{2\omega_0^2}{\omega^2}\Big[-(3t_1+6t_2)U^A_{m,n,X} + t_1\left(U^B_{m,n,X} + U^B_{m+1,n-1,X} + U^B_{m,n-1,X}\right)$$
$$+ t_2\left(-U^A_{m,n+1,Y} + U^A_{m,n-1,Y} + U^A_{m+1,n,Y} - U^A_{m-1,n,Y} + U^A_{m-1,n+1,Y} - U^A_{m+1,n-1,Y}\right)\Big]$$

, (3)

where $\omega_0 = 1/\sqrt{LC}$, with $L$ and $C$ as the inductance and capacitance respectively. We let $C_1 = t_1 C$ and $C_2 = t_2 C$ denote the capacitances for NN and NNN connections respectively.

Similarly, we can derive the equations for inductor Y at site A, as well as for the equations for site B:

$$U^A_{m,n,Y} = -\frac{2\omega_0^2}{\omega^2}\Big[-(3t_1+6t_2)U^A_{m,n,Y} + t_1\left(U^B_{m,n,Y} + U^B_{m+1,n-1,Y} + U^B_{m,n-1,Y}\right)$$
$$+ t_2\left(U^A_{m,n+1,X} - U^A_{m,n-1,X} - U^A_{m+1,n,X} + U^A_{m-1,n,X} - U^A_{m-1,n+1,X} + U^A_{m+1,n-1,X}\right)\Big]$$

,

$$U^B_{m,n,X} = -\frac{2\omega_0^2}{\omega^2}\Big[-(3t_1+6t_2)U^B_{m,n,X} + t_1\left(U^A_{m,n,X} + U^A_{m+1,n-1,X} + U^A_{m,n-1,X}\right)$$
$$+t_2\left(-U^B_{m,n+1,Y} + U^B_{m,n-1,Y} + U^B_{m+1,n,Y} - U^B_{m-1,n,Y} + U^B_{m-1,n+1,Y} - U^B_{m+1,n-1,Y}\right)\Big],$$

$$U^B_{m,n,Y} = -\frac{2\omega_0^2}{\omega^2}\Big[-(3t_1+6t_2)U^B_{m,n,Y} + t_1\left(U^A_{m,n,Y} + U^A_{m+1,n-1,Y} + U^A_{m,n-1,Y}\right)$$
$$+t_2\left(U^B_{m,n+1,X} - U^B_{m,n-1,X} - U^B_{m+1,n,X} + U^B_{m-1,n,X} - U^B_{m-1,n+1,X} + U^B_{m+1,n-1,X}\right)\Big]. \quad (4)$$

Defining $U_{\uparrow,\downarrow} = U_X \pm iU_Y$, the above equations simplify to

$$\left(3t_1+6t_2 - \frac{2\omega_0^2}{\omega^2}\right)U^A_{m,n,\uparrow} = t_1\left(U^B_{m,n,\uparrow} + U^B_{m+1,n-1,\uparrow} + U^B_{m,n-1,\uparrow}\right)$$
$$+t_2\left(e^{i\phi}U^A_{m,n+1,\uparrow} + e^{-i\phi}U^A_{m,n-1,\uparrow} + e^{-i\phi}U^A_{m+1,n,\uparrow} + e^{i\phi}U^A_{m-1,n,\uparrow} + e^{-i\phi}U^A_{m-1,n+1,\uparrow} + e^{i\phi}U^A_{m+1,n-1,\uparrow}\right),$$

$$\left(3t_1+6t_2 - \frac{2\omega_0^2}{\omega^2}\right)U^B_{m,n,\uparrow} = t_1\left(U^A_{m,n,\uparrow} + U^A_{m+1,n-1,\uparrow} + U^A_{m,n-1,\uparrow}\right)$$
$$+t_2\left(e^{i\phi}U^B_{m,n+1,\uparrow} + e^{-i\phi}U^B_{m,n-1,\uparrow} + e^{-i\phi}U^B_{m+1,n,\uparrow} + e^{i\phi}U^B_{m-1,n,\uparrow} + e^{-i\phi}U^B_{m-1,n+1,\uparrow} + e^{i\phi}U^B_{m+1,n-1,\uparrow}\right),$$

$$\left(3t_1+6t_2 - \frac{2\omega_0^2}{\omega^2}\right)U^A_{m,n,\downarrow} = t_1\left(U^B_{m,n,\downarrow} + U^B_{m+1,n-1,\downarrow} + U^B_{m,n-1,\downarrow}\right)$$
$$+t_2\left(e^{-i\phi}U^A_{m,n+1,\uparrow} + e^{i\phi}U^A_{m,n-1,\uparrow} + e^{i\phi}U^A_{m+1,n,\uparrow} + e^{-i\phi}U^A_{m-1,n,\uparrow} + e^{i\phi}U^A_{m-1,n+1,\uparrow} + e^{-i\phi}U^A_{m+1,n-1,\uparrow}\right),$$

$$\left(3t_1+6t_2 - \frac{2\omega_0^2}{\omega^2}\right)U^B_{m,n,\downarrow} = t_1\left(U^A_{m,n,\downarrow} + U^A_{m+1,n-1,\downarrow} + U^A_{m,n-1,\downarrow}\right)$$
$$+t_2\left(e^{-i\phi}U^A_{m,n+1,\uparrow} + e^{i\phi}U^A_{m,n-1,\uparrow} + e^{i\phi}U^A_{m+1,n,\uparrow} + e^{-i\phi}U^A_{m-1,n,\uparrow} + e^{i\phi}U^A_{m-1,n+1,\uparrow} + e^{-i\phi}U^A_{m+1,n-1,\uparrow}\right). \quad (5)$$

Applying a spatial Fourier transform, we obtain eigenvalue equations for spin up and spin down states, parameterized by the wavevector $k$:

$$E\begin{pmatrix}U^A_{k,\uparrow}\\U^B_{k,\uparrow}\end{pmatrix} = \begin{pmatrix}P_k(\phi) & T_k \\ T_k^* & P_k(-\phi)\end{pmatrix}\begin{pmatrix}U^A_{k,\uparrow}\\U^B_{k,\uparrow}\end{pmatrix},$$

$$E\begin{pmatrix}U^A_{k,\downarrow}\\U^B_{k,\downarrow}\end{pmatrix} = \begin{pmatrix}P_k(-\phi) & T_k \\ T_k^* & P_k(\phi)\end{pmatrix}\begin{pmatrix}U^A_{k,\downarrow}\\U^B_{k,\downarrow}\end{pmatrix}, \quad (6)$$

where

$$E = 3t_1 + 6t_2 - 2\frac{\omega_0^2}{\omega^2}, \quad (7)$$

$$T_k = t_1\left(e^{i k \vec{e}_1} + e^{i k \vec{e}_2} + e^{i k \vec{e}_3}\right), \quad (8)$$

$$P_k = 2t_2\left[\cos(k\vec{v}_1 + \phi) + \cos(k\vec{v}_2 + \phi) + \cos(k\vec{v}_3 + \phi)\right]. \quad (9)$$

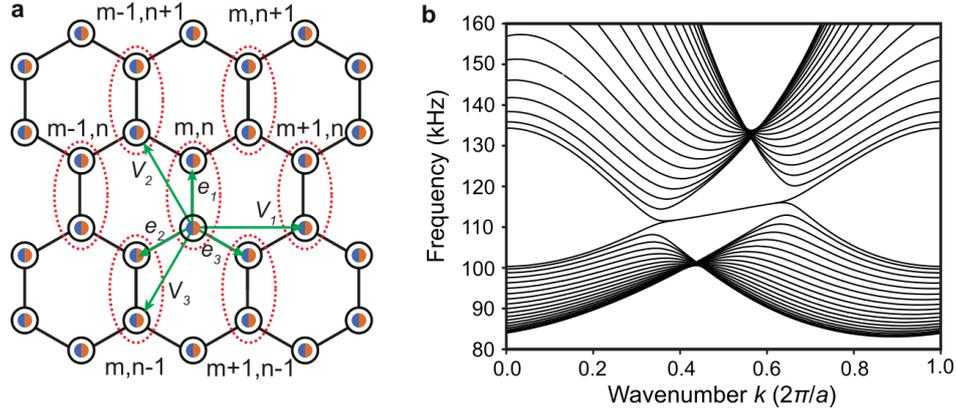

**Fig. S1. Schematic and band diagram of the lattice. a** Schematic of the honeycomb lattice showing the (*m*, *n*) site labels. **b** Band structure for spin-down circuit eigenstates. All other parameters are the same as in Fig. 1**c** of the main text.

The matrices in Eq. (6) have the same form as the Hamiltonian in the modified Haldane model. Thanks to the braided capacitive NNN couplings, the circuit can be regarded as two copies of the modified Haldane model, one for each "spin", and each capable of hosting antichiral edge states. The band diagram for the spin-up states of the circuit, in a strip configuration, is shown in Fig. 1**c** of the main text. In Fig. S1**b**, we show the corresponding band diagram for the spin-down states. Note that the antichiral edge states in the two spin sectors have precisely opposite group velocities.

The band diagrams for different NNN coupling phases are shown in Fig. S2. The NNN coupling phases slightly affect the shape and slope of the antichiral edge state dispersion relation. The group velocities for $\phi = \pi/3$ and $2\pi/3$ are not too different from that of the $\phi = \pi/2$ case (which we used in our experiments). For $\phi = \pi$, the dispersion is flat, which is undesirable for observing the states experimentally.

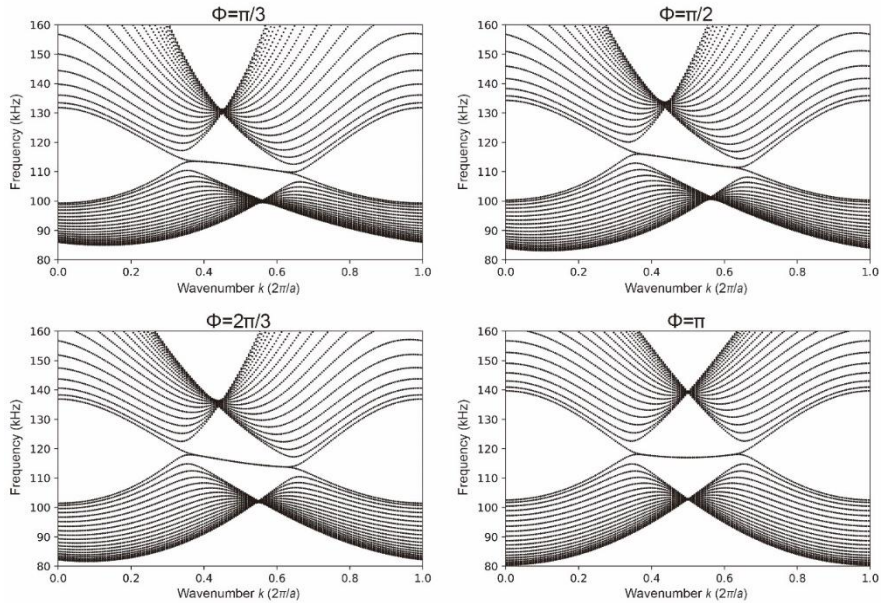

**Fig. S2. Band diagram of the lattice.** Band structure for different NNN coupling phases. All other parameters are the same as in Fig. 1**c** of the main text.

# Circuit Simulations

All circuit simulations are performed using LTspice (www.linear.com/LTspice).

In Fig. S3**a**, we plot the simulation results for steady-state voltage amplitudes in the 128-site sample, corresponding to the experimental results shown in Fig. 3**b** of the main text. The AC (steady-state) spin-up voltage source is located at the upper-rightmost corner site. Similar to the experimental results, a large voltage amplitude is observed on both edges.

The group velocities of the edge states are extracted using dwell time estimates. The same procedure is used in the experiment and simulations. For example, Fig. S3**b** plots the phase $\varphi$ of the spin-up voltage signals versus frequency $f$ for two representative sites. We pick five frequencies within the frequency range in which there is a strong edge state response (as previously noted, this frequency range differs slightly between experiment and simulations), and take the average of the four estimates of the dwell time $d\varphi/d(2\pi f)$. The group velocity is then estimated from the rate of change of dwell time with distance along the edge.

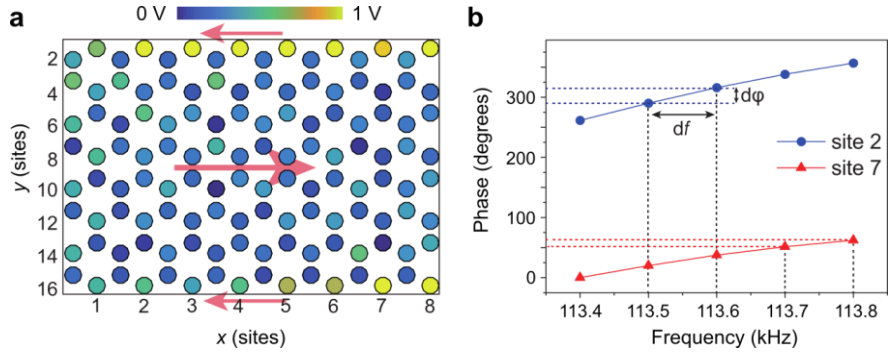

**Fig. S3. Circuit simulations of antichiral edge state propagation in a finite lattice. a** Circuit simulation results showing the heat map of voltage amplitudes. The circuit configuration is the same as in Fig. 3 of the main text. Voltage sources with 90° phase difference (so as to excite spin-up states) are placed on the inductors X and Y at the upper-rightmost corner site. **b** Frequency dependence of the phase of the spin-up voltage response on two typical sites (at $x = 2$ and $x = 7$ on the top edge). This is used to estimate the site-dependent dwell time, and hence the group velocity of the edge state.

To illustrate the dynamics of the lattice, we performed time-domain simulations of a circuit structure with 10×20 sites, using a Gaussian input signal with time delay $t_0 = 50$ μs, width $dt = 50$ μs, and frequency $f = 113.64$ kHz, as shown in Fig. S4**a**. The X and Y inductors at the upper-rightmost corner site are driven by voltages $S_X = e^{-\frac{(t-t_0)^2}{dt^2}} \cos(\omega t)$ and $S_Y = e^{-\frac{(t-t_0)^2}{dt^2}} \sin(\omega t)$, respectively, so as to produce a spin-up source $S_\uparrow = e^{-\frac{(t-t_0)^2}{dt^2}} e^{i\omega t}$. As shown in the first row of Fig. S4**b**, the source excites edge states that propagate leftward along the top edge. Upon reaching the left boundary at around 250 μs, the edge states couple to bulk states and reflect back to the right. At around 450 μs, the edge state at the bottom edge is excited and propagates to the left.

To obtain an independent estimate of the group velocity, we performed another set of time-domain simulations with a longer lattice site of 40 sites (rather than 10 sites). With a frequency width of 10 kHz for the input signal, we identify the peak positions of the left-moving pulse along the edge at different elapsed times. From these, we obtain an estimated group velocity of -0.057 sites/μs, which agrees very well with the group velocity of -0.059 sites/μs from the band diagram obtained via frequency domain simulations.

For the sake of clarity, in these time-domain simulations the resistances in the inductors have been omitted, so as to suppress the decay rate of the edge and bulk states (these resistances are included in the previously-described steady-state circuit simulations).

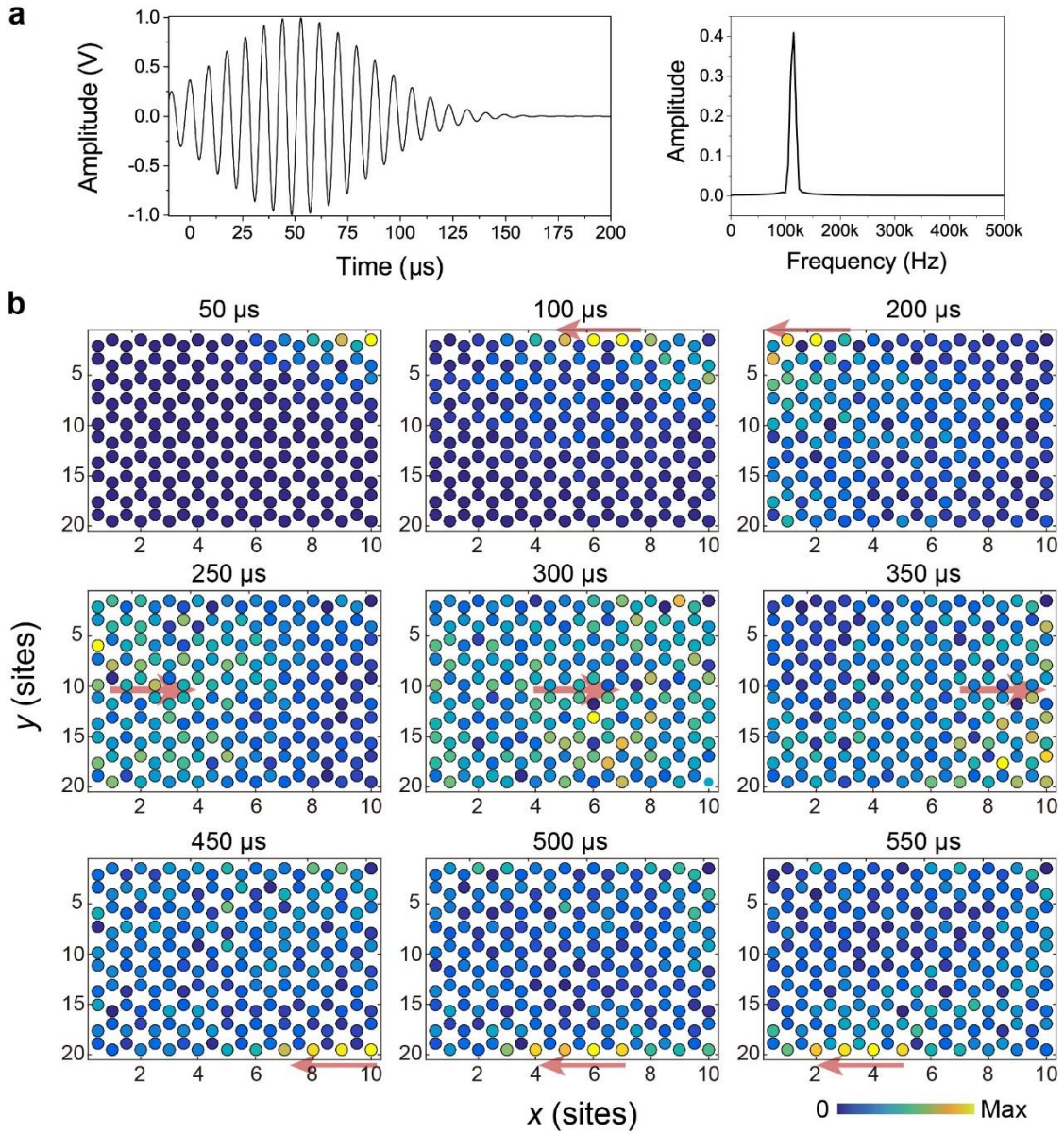

**Fig. S4. Time-domain circuit simulation results. a**, Time dependence of the Gaussian voltage

signal injected into the circuit lattice (left panel), and the frequency spectrum (right panel). The circuit configuration, including the location of the sources, is the same as in Fig. 3 of the main text. **b**, Heat maps of voltage amplitude distributions at different times, showing the propagation of the left-moving edge state on the top edge, reflection into bulk states (around 200 μs), and subsequent reflection into left-moving edge states on the bottom edge (around 450 μs). The directions of propagation are indicated by red arrows.

## Möbius strip geometry

A Möbius strip configuration can be realized by connecting the left and right boundaries of a rectangular sample with a "180 degree twist". For example, consider the 10×20 sample shown in Fig. S5. Site 1 on the left boundary (at the upper-right corner) is connected to site 20 on the right boundary (at the lower-right corner), and so forth.

With such connections, consider an edge state moving leftward along the top edge. Upon reaching site 1 on the left boundary, it should continue to site 20 on the right boundary, and continue moving leftward along the bottom edge. This is not consistent with the chiral nature of the edge states of an unpaired Chern insulator. In a circuit lattice that implements two copies of a Chern insulator, this means that edge state must flip into the opposite spin when it crosses to the other edge.

In the modified Haldane lattice, however, the Möbius boundary conditions do not induce spin-flip. As the time-domain circuit simulations of Fig. S5 show, an antichiral edge state retains its spin and continues propagating in the same direction when it crosses the boundary onto the opposite edge.

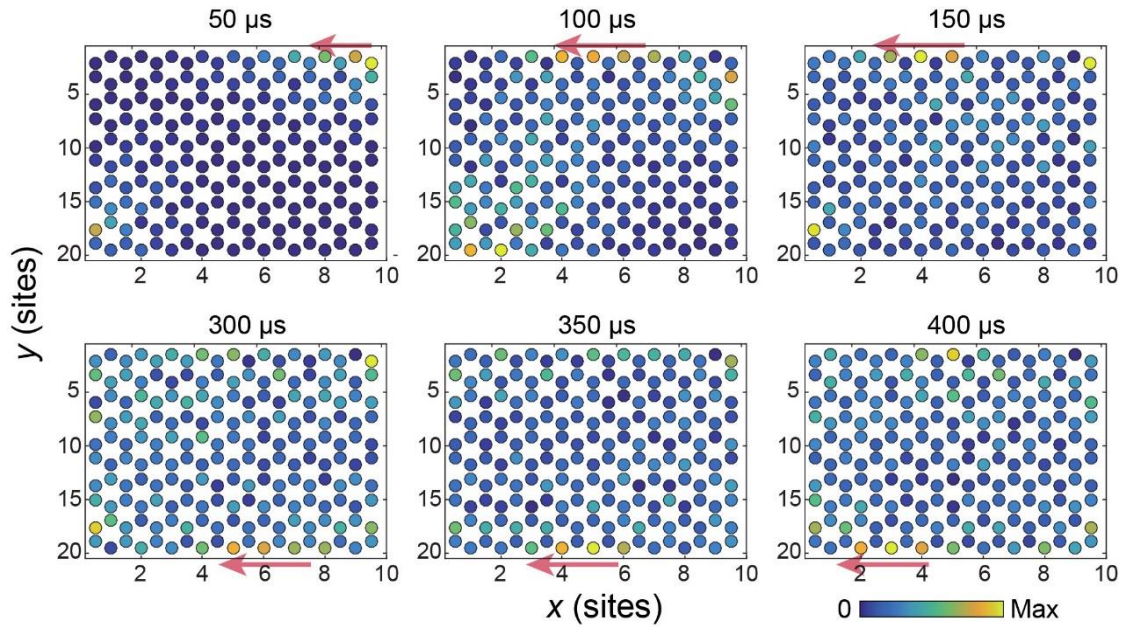

**Fig. S5. Time-domain circuit simulations with a Möbius strip configuration.** A spin-up source, located at the upper-rightmost site, excites an antichiral edge state that moves leftward along the edge. At around 300 μs, it reaches the left boundary, switches to the bottom edge (thanks to the Möbius boundary conditions), and continues propagating leftward.